# Determinants Influencing Intention to Use Social Commerce for Shopping in developing countries: A Case Study of Oman

S. Al Harizi, M. Al Areimi, A. K. Shaikh

Department of Information systems, Sultan Qaboos University, Muscat, Oman,
s128580@student.squ.edu.om, s18284@student.squ.edu.om, shaikh@squ.edu.om[*]

**Abstract –** Social media has had a significant impact on our individual lives, including our behavior regarding the purchasing of daily products. This study investigates the factors influencing Omani nationals' intentions to obtain products via social commerce. The researcher surveyed 202 participants and utilized the Technology Acceptance Model to develop the theoretical framework. The data collection was analyzed statistically using an appropriate testing mechanism. Statistical methods, including Cronbach's alpha and multiple linear regression, were utilized for reliability and hypotheses testing. After analyzing the collected data and testing the hypotheses, the findings indicated that perceived usefulness, enjoyment, and ease of use of social commerce affect positively on Omani nationals' intentions to utilize social commerce for shopping. The independent variables had a statistically significant impact on the intention to use social commerce shopping for products; these explain 69.9% of the variation on customers' intention to utilize social commerce for shopping.

**Keywords:** Social Commerce, Technology Acceptance Model, TAM, Linear Regression, Online Shopping, Oman.

[*] Corresponding Author : A. K. Shaikh, shaikh@squ.edu.om

# Introduction

The growth of the internet and social media has had a significant impact on our individual lives as it affects the way we communicate, learn, think, and shop. Due to the rising importance of social networking platforms, there is increased interest and significance in the digital world's advancement, especially around recent developments in the field of data processing. Social networks, often identified as online social networks or social networking platforms, are an essential part of the confluence between the real and the virtual world in Web 2.0. Web 2.0 technology allows sharing and collaboration to internet users and help them to express themselves online. The technology offers the ground to obtain, understand, and sustain social interrelations with other users who have comparable interests [1]. Nowadays, there are numbers of social network sites and platforms. Facebook is the largest social network site in the world, but other applications such as WhatsApp and Instagram are also widely popular [2]. Social commerce sites (SCSs) provide online purchasing and selling activities launched on social media. They offer commercial transactions via social media or other e-commerce websites [3]. SCSs incorporate social media functionalities that reach the advancement of individuals' relationships and relationships between groups of consumers [4]. In recent years, the success and growth of social commerce have been driven by the exchange of knowledge about consumer products. According to the research study by [5], the best way to perform social commerce is when company strategy aligned with the strategy of social media. This advancement in social commerce is dependent upon two key determinants: customer purchasing power and rapid advancement in information technology, especially Web 2.0 [3]. New internet technologies such as Web 2.0 affect business development by expanding the reach and reducing time to market [6]. In addition, this trend transforms the relationship between businesses and their customers fundamentally and drives them to find new and innovative ways of reaching and interacting with their customers. Social networks can play a critical role in the future of marketing because they can externally substitute consumer frustration with interaction. While internally, they turn the traditional control focus transparently and collaboratively, which is more beneficial to success in the new business environment [6].

Social commerce in Oman constitutes a sizeable commercial market. According to a report from the Omani National Center for Statistics and Information [7], One-third of Omanis have accessed social networking sites or applications to purchase or look for certain products and services. Also, one in five Omanis has utilized social media for business purposes, such as advertising or selling specific products or services. Social media use for business purposes is substantially higher among females (40%) than males (23%) when it comes to buying or looking up products and services [7].

This paper adopted the Technology Acceptance Model (TAM) for the development of a theoretical framework. There are different adoption models such as TAM, TAM 2,

TAM 3, UTAT and UTAT 2; however, TAM was selected because it provides an informative representation that lays out choices that influence the acceptance of users. Therefore, TAM is helpful for evaluating and predicting the level of user readiness to accept using information technology [8]. The TAM model states that the behaviors affecting an individual's acceptance of information technology are defined by four factors: perceived ease of use, usefulness, enjoyment, and intentions to use social commerce. In a recent research study [2] argue that TAM is a significant predictor of technology acceptance. Drawing on the TAM, this study aimed to investigate the impacts of perceived ease of use, usefulness, and enjoyment of social commerce on the intentions of consumers to use social commerce when purchasing products in Oman. This study broadens research on the online retail business and addresses the following research questions related to determinants affecting intentions to use social commerce:
  A. Does perceived ease of use of social commerce influence social commerce intention?
  B. Does perceived usefulness of social commerce influence social commerce intention?
  C. Does perceived enjoyment of social commerce influence social commerce intention?

The remainder of this paper is organized as follows: Section 2 reviews the existing literature. The research methodology is presented in Section 3. Section 4 presents results and discussion. Section 5 outlines the recommendations, and section 6 concludes this study and outlines implications for future work.

## Literature Review

The existing research studies have examined consumer intentions in dealing with online and social commerce. Some recent studies have examined the factors that contribute and facilitate the adoption of social commerce to online purchase intention among Malaysian consumers [9]. The sample of this paper was limited only to Malaysian university student. While our study will fill this gap of different age range with different education level not limited to university students. Also, [9] has investigated the relationship between the factors and mentioned for further study to explore the effects of the ease of use on the intention to use social commerce. [10] examined only two factors from the TAM model and in the Dhofar region where our research will examine more factors and in the different governorate. Further investigation is required to understand the factors related to customer intentions to use social commerce for shopping, as the use of social media increases rapidly.

*Technology Acceptance Model*

Technology Acceptance Model (TAM) introduced in 1986 by Fred Davis [11], [12]. It is coming to replace the attitude measure of the theory of reasoned action (TRA) with technology acceptance measures. It is the most widely applied model in the information system (IS) field. The most important purpose of this model is to predict the acceptability of IS. It aims to predict the acceptability of tools that could be brought to the system in order to make them acceptable to users. It focuses on cognitive factors to predict technology acceptance. This model compromise of two main factors to determine IS which are: perceived usefulness and perceived ease of use. Also, [13] found that variables from UTAUT have a strong correlation with Social Commerce behavior.

Perceived usefulness defines the degree of person believes that the use of the system will assist them to improve their performance. However, perceived ease of use based on the degree person believes that the system use will require less effort. The model has been used for several types of research in providing empirical evidence between usefulness and ease of system use. TAM theories assume that the IS determined by the behavioral intention that determined by personal attitude toward the system use and its utility from their perception. Although, the individual attitude cannot be determined by only one factor which is the system use but can also depend on the system impact upon individual performance.

The research study [14] has added the perceived enjoyment factor to the TAM, which defines the degree of enjoyment from an individual's perception in using systems or technology [15]. So, TAM theory framework had suggested three factors that could affect an individual's intention from three previsions which are ease of use, usefulness, and enjoyment [16], [17]. The relationships between perceived usefulness and perceived ease of use under the context of specific information systems have been explored by the research study [18]. In that research, TAM model was well presented throughout the study that provided a meaningful reference of technology for future researcher and scientist who may plan to improve the acceptance and promote the innovation of technology.

A popular study by [19] predicted consumer acceptance of e-commerce by using TAM model and the findings of the study showed that the perceived usefulness and ease of use have a significant effect on transaction intentions of e-commerce. Furthermore, findings validate the conceptualization for the imperative role of trust in e-commerce and explicitly describes its precise effects.

Since shopping via social commerce platform depending on some factors affecting customer intention. Some customers perceive ease of use as factors for shopping using social commerce, while others perceive usefulness as the main factor to make order according to social commerce. However, some customer depending on their enjoyment felling to buy through social commerce. So, TAM serves as a useful framework, and it is consistent with several investigations into the determinants that influence customer intention for shopping using social commerce.

Determinants affecting using social commerce for shopping:

### A. Perceived Ease of Use (PEOU)

The Technology Acceptance Model (TAM) determines how a specific technology will be accepted. It is based on individual belief if this technology free from difficulties and did not require great effort to use and understand it. It is accepted in a wide range of studies that suggest that the possibility of customers to make use of new technology will grow if the technology is considered convenient easy to use and useful [20].

Ease of use is examining the relationship between perceived ease of use and attitudes toward online commerce founded in previous studies, for example, the research study about factors affecting social commerce acceptance in Lithuania that was done by [21]. They came up with questionnaires that show the impact of ease of use in social commerce in the country Lithuania and their result show that ease of use has the lowest impact on an individual's intention in using social e-commerce.

Another research that aims to represent the impact of using social commerce in fashion product conducted by [22] that show there are indirect and insignificant impact between customer intention and ease of use. However, [23] in his research about determinants that influence a consumer to use online shopping, the author reached to that ease of use has both direct and indirect impact based on consumer background and education level. There is a positive relationship between ease of use and usefulness, so if online shopping is easy to use, it will help the consumer to find the usefulness as proved in the same time [24] study.

### B. Perceived Usefulness (PU)

Perceived usefulness is one of the Technology Acceptance Model determinants that determine how a specific technology will be usefulness. It is based on individual belief if this technology will help to improve performance. Perceived usefulness has a positive effect on consumers in Oman to shift from traditional stores to online buying [25].

Usefulness has a greater and strong impact on the social commerce of fashion product than ease of use and enjoyment based on result presented by [22]. However, in [23] research shows that the usefulness has an indirect impact on consumer intention on online shopping. On the other hand, [24] confirm that usefulness directly influences the online purchase intention using social commerce.

### C. Perceived Enjoyment (PE)

It is another factor of Technology Acceptance Model that determines how a specific technology will be triggered enjoyment. It is based on individual belief if this technology interest.

Many studies emphasize that ease of use and usefulness have positively influence perceive of enjoyment as mentioned by [22]. Moreover, according to the research result of [26] consumers were more interested in the enjoyment factor that has a direct impact to improve and continue using social commerce sites for purchasing. Similarly, to study conducted by [27] that approve enjoyment perception has a stronger impact on attitudes towards social commerce in apparel shopping.

*Research Hypotheses*
H1: Perceived ease of use positively affects intentions to use social commerce shopping.
H2: Perceived usefulness positively affects intentions to use social commerce shopping.
H3: Perceived enjoyment positively affects intentions to use social commerce shopping.

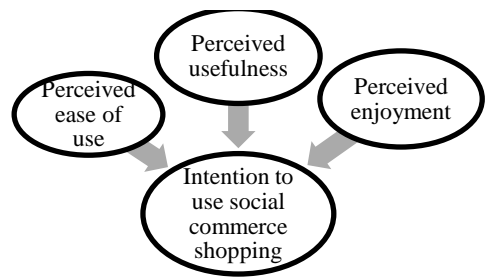

**Fig. 1** Research Model

## Research Methodology

As this study has to quantify attitudes, opinions and behaviors of consumers in the context of intentions to use social commerce when shopping for products, quantitative research is used to investigate three hypotheses to the influence of consumers' perceived ease of use of social commerce, usefulness, and enjoyment on the consumers' intention. The following sections describe the sampling, survey instrument development, data collection procedure, and data analyses.

### A. Population and Sample

The study sample was from the population of three towns in Oman such as Muscat, Salalah and Sur to investigate the nationals' intention to use social commerce for shopping. The collection of data was conducted via an online questionnaire (Google Form) by having different Omani respondents. Based on this questionnaire we reached around 202 respondents from the three towns in Oman.

### B. Survey instrument

A survey was distributed online to test the suggested model as mentioned in **Fig. 1**

Reliable and accurate scale objects were adapted from prior studies and revised to relate to the issue of social commerce, as given in (Appendix 2).

### C. Collection of Data

The collection of data was through online questionnaires. The structured questionnaires were composed in English that prepared based on specific research purposes. A quantitative approach was used to process data collection.

### D. Methods of Data Analysis

Data analysis was used with a suitable statistical test to analyze the collected data. In this research different statistical methods were used to examine the reliability test Cronbach's alpha and hypotheses test like simple and multiple linear regression via SPSS software.

## Results and Discussion

The reliability of the study and Cronbach's alpha are conducted to ensure that the results are reliable as shown in **Table 1** that shows represent the Cronbach's alpha of all variables are more than 0.6 which are considered reliable [28]. **Table 1** indicates that the alpha Cronbach's coefficient which is ranging from (0.778) for perceived ease of use, (0.904) for the perceived usefulness, (0.932) for the perceived enjoyment, and (0.954) for customers intending to use social commerce for shopping. Further, a high level of study variables availability reflected through the mean.

### A. Descriptive Analysis

**Table 1** Descriptive and Reliability Statistics

|  | N | Mean | Std. Deviation | Cronbach's Alpha |
|---|---|---|---|---|
| PEU | 202 | 3.662 | .810 | .778 |
| PU | 202 | 3.736 | .872 | .904 |
| PE | 202 | 3.734 | .936 | .932 |
| ITU | 202 | 3.711 | .988 | .954 |
| Valid N (listwise) | 202 | | | |

**Table 2** Frequency Calculation

|  |  | Frequency | Percent (%) |
|---|---|---|---|
| Gender | Male | 56 | 27.7 |
|  | Female | 145 | 71.8 |
| Age | 18 or less | 6 | 3.0 |
|  | 19 – 30 | 95 | 47.0 |
|  | 31 – 45 | 85 | 42.1 |
|  | 46 and above | 16 | 7.9 |
| | Less than Secondary School | 2 | 1.0 |
| | Secondary School | 22 | 10.9 |
| Education | Diploma | 35 | 17.3 |
| | Bachelor | 116 | 57.4 |
| | Master and PhD | 27 | 13.4 |
| | 0 – 500 OMR | 88 | 43.6 |
| Salary | 501 – 1000 OMR | 47 | 23.3 |
| | 1001 – 1500 | 35 | 17.3 |
| | 1501 and above | 19 | 9.4 |

**Table 3** Frequency of Websites Use

| Websites | Frequency |
|---|---|
| Jollychic | 60 |
| AliExpress | 51 |
| Boutiqaat | 25 |
| eBay | 16 |
| Namshi | 39 |
| Amazon | 46 |
| WhatsApp | 92 |
| Instagram | 113 |
| Other | 44 |

### B. Hypotheses Testing

#### 1. Simple Linear Regression

First hypothesis: testing of the first hypothesis, the result was as the following:

**Table 4** Regression Analysis Results

| R Square | F | Sig. |
|---|---|---|
| .375 | 119.939 | .000[b] |

a. Dependent Variable: ITU
b. Predictors: (Constant), PEU

**Table 5** Hypothesis Test Results

| Model | B | Std. Error | Beta | t | Sig. |
|---|---|---|---|---|---|
| PEU | .747 | .068 | .612 | 10.952 | .000 |

a. Dependent Variable: ITU

Based on the results, it is noticed that the perceived ease of use which is appeared in **Table 5** has a direct impact ($\beta = 0.612$) on the intention to use social commerce shopping. From **Table 4**, we found that ($R^2 = 0.375$) the meaning of this is that perceived ease of use has been explained (37.5%) of the variance in on intention to use social commerce shopping. Also, we noticed that from **Table 4** the statistical value which is (F = 119.939) is statistically important at (0.000 < 0.05). This result contradicts with others the results like what done by [22] that show there is an indirect and insignificant impact between customer intention and ease of use and with [21] that reach to that ease of use has the lowest impact on individual's intention in using social e-commerce in Lithuania. Also, similar findings were reported by [29]. This result clarifies that most social commerce ignores the ease-of-use determinants, that has negative impacts on consumers intention to use online shopping. So, as advice buyers through social commerce should give ease of use a significant interest to gain more consumers.

Second hypothesis: testing of hypothesis, the result is as follows:

**Table 6** Regression Analysis Results

| R Square | F | Sig. |
|---|---|---|
| .484 | 187.867 | .000[b] |

a. Dependent Variable: ITU
b. Predictors: (Constant), PU

**Table 7** Hypothesis Test Results

| Model | B | Std. Error | Beta | t | Sig. |
|---|---|---|---|---|---|
| PU | .788 | .058 | .696 | 13.706 | .000 |

a. Dependent Variable: ITU

 Based on the analysis results, (perceived usefulness) as it is represented from **Table 7** has a direct effect ($\beta = 0.696$) on the intention to use social commerce shopping. Further, as it is seen from **Table 6** ($R^2 = 0.484$) explains that the perceived usefulness has been explained (48.4%) of the change in intention to use social commerce shopping. Also, we found that from **Table 6** the statistical value ($F = 187.87$) was significant statistical at ($0.000 < 0.05$).

Third hypothesis: testing the hypothesis that results as follow:

**Table 8** Regression Analysis Results

| R Square | F | Sig. |
|---|---|---|
| .632 | 343.869 | .000[b] |

a. Dependent Variable: ITU
b. Predictors: (Constant), PE

**Table 9** Hypothesis Test Results

| Model | B | Std. Error | Beta | t | Sig. |
|---|---|---|---|---|---|
| PE | .839 | .045 | .795 | 18.544 | .000 |

a. Dependent Variable: ITU

 Based on the regression analysis results, we found that the perceived enjoyment (X3) as it seen in **Table 9** has a direct impact ($\beta = 0.795$) on the buyers' intention to use social commerce shopping (Y). As displayed in **Table 8**, ($R^2 = 0.632$) which means that perceived enjoyment has been explained (63.2%) of the change in intention to use social commerce shopping. Further, we found that in **Table 8** that the statistical test value ($F = 343.87$) is a significant statistical at ($0.000 < 0.05$). This result is consistent with the study conducted by [27] which approve an enjoyment perception that has a stronger impact on attitudes towards social commerce in apparel shopping.

### 2. *Multiple Regression*

This method used to test all independent variables on the dependent variable, the results are mentioned as below:

**Table 10** Regression Analysis Results

| R Square | F | Sig. |
|---|---|---|
| .699 | 153.556 | .000[b] |

a. Dependent Variable: ITU
b. Predictors: (Constant), PE, PEU, PU

**Table 11** Hypothesis Test Results

| Model | B | Std. Error | Beta | t | Sig. |
|---|---|---|---|---|---|
| PEU | .204 | .061 | .167 | 3.346 | .001 |
| PU | .272 | .062 | .240 | 4.358 | .000 |
| PE | .571 | .058 | .541 | 9.877 | .000 |

a. Dependent Variable: ITU

Table 10 shows that ($R^2$ = 0.699) which explain that the perceived ease of use, perceived usefulness and perceived enjoyment have been explained (69.9%) of the change on the intention to use social commerce shopping for products. Moreover, we noticed that in Table 10 the statistical test value (F = 153.56) is significant statistically at (0.000< 0.05). Based on the multiple regression analysis results, we found that in Table 11 the perceived ease of use ($\beta$ = 0.167, sig. = 0.001), perceived usefulness ($\beta$ = 0.240, sig. = 0.000), and perceived enjoyment ($\beta$ = 0.541, sig. = 0.000) have a direct effect on the intention to use social commerce shopping. All constructs perceived ease of use (t = 3.35), perceived usefulness (t = 4.36), and perceived enjoyment (t = 9.88) were significant.

A study from Jordan [30] found that Perceived Ease of Use and Perceived Usefulness have no direct impact on consumers' attitude, but have an indirect impact on intention through attitude. Similar results founded the US results revealed that ease of use and usefulness has an indirect effect on attitudes towards social commerce [27]. However, a study conducted in Sri Lanka was revealed that online purchase intention positively and significantly related to perceived usefulness and perceived ease of use [24] which is in line with the current study. According to a paper conducted in the US found that perceived enjoyment has the strongest impact on attitudes, and attitudes have a strong impact on consumer intentions for adopting social commerce for apparel purchases [27].

## Recommendations

Our findings suggest that social commerce sellers should make the customers feel connected to these platforms and users on social media. Also, marketers would offer a range of content that is up to date on their social commerce platforms where the consumers feel enjoyable and interested by sharing with other users that help them for buying decision. Based on the results, these website sellers should increase their availability and usability by designing the experiences of social commerce to be more useful, easy to use as well as enjoyable for the buyers to get access to experience features of such products or have information on the product. For example, some of the marketers use Instagram to shop with photo tags that offer in-app details about particular prices, features, and descriptions of the product which is a button with the title "Shop Now" [22]. A few days ago, Instagram has launched a new feature for business accounts to advertise their products as a separate story called "Support Small Business". These tools assist both parties in which customers can easily reach the preferred products and sellers can reach many numbers of new customers. Thus, the study results supported the importance of the experiential and informative element on social commerce means to improve intentions' behavior between customers [31].

## Conclusion & Implications

This research aimed to comprehend the influences affecting the intentions of consumers who use social commerce for shopping. The study focused on the investigation of consumers in relation to several factors that could influence their intentions on using social commerce for shopping, namely ease of use, usefulness, and enjoyment. In order to study these factors, this research relied on the Technology Acceptance Model (TAM) theory to provide a framework that supported the purpose of this study. The study's findings confirm that all variables – perceived usefulness, perceived ease of use, and perceived enjoyment – are positively influenced by customers' intention to purchase products from social commerce. The results of the study, therefore, support all the hypotheses that are helpful for stakeholders who use social commerce for business activities, to take all these perceive in their point of view to guarantee their business continuity and attract more consumers. By tracking customers' behavior electronically, sellers can learn more about them and take advantage of the biggest impacts that influence their buying intentions and relationships with the help of semantic technology [32] and social network analysis functions [33]. Future work is needed to explore the impact of these technologies on consumer behavior.

This study has generated theoretical and practical implications of social commerce for shopping and has provided some recommendations. In the theoretical aspect, this research has provided a framework that describes the effect of social commerce factors on buyers' intentions to use online shopping. Moreover, this study explored multiple factors that mixed variables from two models adjusted to the research setting in Oman. While this study is expected for further discussion in which can be tested and developed in different settings. The findings suggest the need for further research in which these variables can be explored and tested in different settings. In the practical aspect, the implications provide several recommendations for businesses and sellers. Businesses can take advantage of this study due to expected insights related to the social commerce concept in order to gauge the intention of customers for using online shopping. By understanding the concept of social commerce factors, marketers can acquire a wider perspective regarding the customers' intention to buy. Furthermore, these factors can be used to lay out strategies that can improve company sales. Also, sellers can use this research to customize their strategies towards the customers' wants.

Although this study broadens existing research on the online retail business of social commerce, some limitations need to be acknowledged. Firstly, the sample was limited to a subset of Omani nationals; thus, the results may not represent the Oman population in general. Because the majority of the sample consisted of Omani citizens located in three towns in Oman, the findings may not correspond with individuals from different backgrounds and other particular regions of Oman. Expanding the research to other areas of Oman would add considerably to understanding consumer intentions when utilizing social commerce for shopping to buy or sell products. Second, further study is needed to examine the relations between perceived ease of use, usefulness, enjoyment, and intentions to use social commerce for shopping. Third, an additional examination is required to investigate other factors (e.g., perceived value and perceived risks) that impact goods purchasing via social commerce.

# Biography

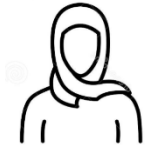

**Shamma Al Harizi** is a Master's student with the Department of Information Systems at Sultan Qaboos University. She obtained her Bachelor's degree from Dhofar University in 2019. She has good skills in scientific research and statistical analysis in the field of Information Systems. Her research interests include E-commerce and Social Commerce.

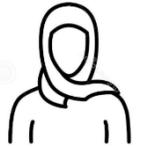

**Maryam Al Araimi** is working as a Computer Programmer at Consumer Protection Authority in Oman. I hold my Bachelor's Degree in Information System from Sultan Qaboos University. Currently, I am studying for a Master's Degree in Information System at Sultan Qaboos University. And I am working on my master's thesis, which focuses on smart city indicators in Oman. My research interest includes Smart Cities, AI, IOT, Social Commerce and Cyber Security.

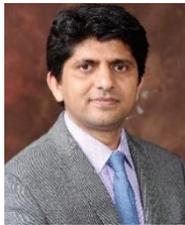

**Abdul Khalique Shaikh is** an Assistant Professor associated with the Department of Information Systems at Sultan Qaboos University Muscat Oman. He has received his PhD degree from a highly reputable Australian University Monash in 2013 and Master degree from University of Detroit USA with distinction. He has a strong skill set and experience gained from working in both academia and industry that endow him with a broad background and a unique perspective on research & teaching. His research interest includes Social Network Analytics, Big Data Analytics, Data Science, Data Governance, E-participation and Blockchain Technology.

# APPENDIX 1: Survey Instrument

The questions below ask about *your general background information*. Please select the appropriate information.

1. What is your gender?
    1. Male
    2. Female
2. What is your age?
    1. 18 or less
    2. 19 – 30
    3. 31 – 45
    4. 46 and above
3. What is your educational qualification?
    1. Less than Secondary School
    2. Secondary School
    3. Diploma
    4. Bachelor
    5. Master and PhD
4. What is your salary level?
    1. 0 – 500 OMR
    2. 501 – 1000 OMR
    3. 1001 – 1500 OMR
    4. 1501 and above
5. What are the social commerce websites you use for shopping (You can select more than one)?
    1. Jollychic
    2. AliExpress
    3. Boutiqaat
    4. eBay
    5. Namshi
    6. Amazon
    7. Account on WhatsApp
    8. Account on Instagram
    9. Other: _______________________________________________

# APPENDIX 2: Survey Instrument

| # | Variables | Items | Source |
|---|---|---|---|
| 1 | **Perceived Ease of use** | Use of social commerce is clear and understandable. | [34] (Kim, Ma, & Park, 2009) |
| | | Use of social commerce is easy for me to become skillful at shopping for products. | |
| | | Use of social commerce does not require a lot of mental effort. | |
| | | Use of social commerce allows me to shop the way I want to. | |
| | | Use of social commerce is easy to learn. | |
| 2 | **Perceived Usefulness** | Use of social commerce increases my productivity in shopping | [34] (Kim, Ma, & Park, 2009) |
| | | Use of social commerce enhances my shopping experience. | |
| | | Use of social commerce is useful in shopping. | |
| | | Use of social commerce improves my shopping skills. | |
| | | Use of social commerce enhances my effectiveness in shopping. | |
| 3 | **Perceived Enjoyment** | Shopping for products through social commerce is interesting. | (Kim, Ma, & Park, 2009) |
| | | Shopping for products through social commerce is fun. | |
| | | Shopping for products through social commerce is entertaining. | |

|   |   | Shopping for products through social commerce is appealing. |   |
|---|---|---|---|
|   |   | Shopping for products through social commerce is enjoyable. |   |
|   |   | Shopping for products through social commerce is exciting. |   |
| 4 | **Intentions to Use Social Commerce** | I intend to buy products through social commerce. | Lee et al. (2006) |
|   |   | In the future, I would be very likely to shop using social commerce for products. |   |
|   |   | I would be willing to buy products through social commerce. |   |
|   |   | I would be willing to recommend my friends to buy products through social commerce. |   |
|   |   | I would visit social commerce to buy products again. |   |